\documentclass[runningheads]{llncs}

\usepackage{amsthm}
\usepackage[T1]{fontenc}
\usepackage{amsmath}
\usepackage{amsmath, bm}
\usepackage{array}%需要用到该宏包
\usepackage{footnote}
\makesavenoteenv{tabular}
\usepackage{caption}
\usepackage{graphicx}
\usepackage{float}
\usepackage{subfigure}
\usepackage{amsthm,amsmath,amssymb}
\usepackage{mathrsfs}
\usepackage{xcolor}
\usepackage{dutchcal}
\usepackage{pifont}
\usepackage{makecell}
\usepackage{wasysym}

\begin{document}
\title{Eavesdropping Mobile Apps and Actions through Wireless Traffic in the Open World}
\titlerunning{Eavesdropping Mobile Apps and Actions}
\author{Xiaoguang Yang\and
Yong Huang\thanks{*Corresponding author} \and
Junli Guo \and
Dalong Zhang \and
Qingxian Wang}
\authorrunning{X. Yang et al.}
\institute{School of Cyber Science and Engineering, Zhengzhou University, Zhengzhou 450001, China \\
\email{yonghuang@zzu.edu.cn}
}

\maketitle              % typeset the header of the contribution
\begin{abstract}
% The abstract should briefly summarize the contents of the paper in
While smartphones and WiFi networks are bringing many positive changes to people’s lives, they are susceptible to traffic analysis attacks, which infer users’ private information from encrypted traffic. 
Existing traffic analysis attacks mainly target at TCP/IP layers or are limited to the closed-world assumption, where all possible apps and actions have been involved in the model training. 
To overcome these limitations, we propose MACPrint, a novel system that infers mobile apps and in-app actions based on WiFi MAC layer traffic in the open-world setting. 
MACPrint first extracts rich statistical and contextual features of encrypted wireless traffic. 
Then, we develop Label Recorder, an automatic traffic labeling app, to improve labeling accuracy in the training phase. 
Finally, TCN models with OpenMax functions are used to recognize mobile apps and actions in the open world accurately. 
To evaluate our system, we collect MAC layer traffic data over 125 hours from more than 40 apps. 
The experimental results show that MACPrint can achieve an accuracy of over 96\% for recognizing apps and actions in the closed-world setting, and obtains an accuracy of over 86\% in the open-world setting.

\keywords{Wireless security  \and  Mobile App fingerprint \and Wireless traffic analysis.}
\end{abstract}
\section{Introduction}
In the recent decade, smartphones have become an integral part of people's everyday lives. 
It is reported that in 2023, smartphone users averagely check their smartphones 144 times and spend 4 hours and 25 minutes on smartphones daily in the United States~\cite{phoneusage}.
Smartphones have reshaped our society in many aspects, including work, entertainment, and social communication through a variety of apps~\cite{zhu2019towards}.
Nowadays, smartphone users prefer the Internet connection via a WiFi network wherever it is available, such as homes, offices, and hotels for cheap and fast connectivity. 
Although WiFi networks generally adopt encryption standards like WPA2 and WPA3 to secure data confidentiality, they are still susceptible to traffic analysis attacks that infer the apps and even actions from encrypted wireless traffic~\cite{backes2015data,wang2015know}.
Since apps in smartphones carry a considerable amount of personal privacy information~\cite{stober2013you,taylor2017robust,tu2018your}, such as gender, age, and hobbies, it is of great importance to investigate the extent to which smartphones are vulnerable to such attacks~\cite{bozorgi2022still,al2019bimorphing}.

Extensive endeavors have been devoted to traffic analysis attacks on smartphones.
Most existing related work analyzes the side channel information of TCP/IP layer traffic~\cite{aceto2020toward,conti2015can,hussey2022positive,taylor2016appscanner,van2020flowprint,xiang2018appclassifier}, and then uses relevant machine learning classification algorithms to classify encrypted traffic and identify the apps used by users. 
Although these methods can identify apps and even actions, in order to obtain TCP/IP layer traffic, it is often necessary to directly connect to wireless access points, routers, or switches. 
In real-life scenarios, the above conditions are often difficult to achieve due to the existence of physical isolation and other protective measures. 
The recent research focuses on attacking MAC layer traffic~\cite{atkinson2018your,wang2015know,zhang2011inferring}, mostly identifying apps by analyzing statistical features such as frame size, direction, and time intervals. 
These methods work well under the closed-world assumption, where apps presented in the recognition stage must also be present during model training. 
However, there are millions of mobile apps that can run on smartphones, thus we can't build fingerprints for all apps.
So these methods' recognition accuracy greatly decreases in the open world.
Hence, none of the existing approaches simultaneously meet the requirements of identifying apps and actions through analyzing wireless encrypted traffic in the open world.

This paper presents MACPrint, a novel attack system that identifies smartphone apps and actions via encrypted WiFi MAC frames over the air.
Due to the differences in app communication protocols and specifications, users' app and action preferences would trigger different changing patterns in WiFi frames.
In this way, we can infer apps and actions based on different traffic patterns.

\textbf{Challenges.} First, in a WiFi network, transmitted data is encrypted in the payload in an 802.11 data frame, which leaves little information about apps and actions per-formed on users' phones. To deal with this issue, we first segment raw traffic into effective traffic traces and represent them using plaintext metadata. Then, various statistical and contextual features are extracted with different sliding windows to generate app and action feature samples.
Second, most of the existing approaches~\cite{baek2023targeted,li2022foap,ni2023eavesdropping} to fingerprint apps and actions rely on manual annotation for traffic traces and adopt closed-world assumption, which yields poor classification performance in the real world.
To avoid this problem, we leverage temporal convolutional networks with OpenMax functions to facilitate app and action classification in open-world setting.
Moreover, an Android app called the Label Recorder is developed to run on smartphones to facilitate fine-grained sample annotation for classifier training.

\textbf{Contributions.} The main contributions of this work are summarized as follows.
We propose MACPrint, a traffic analysis system that is difficult to detect. 
It can effectively extract fine-grained features of encrypted traffic and build app and action fingerprints through a TCN classification model with an OpenMax layer.
We develop an Android app called the Label Recorder that can automatically perform fine-grained sample annotation during classifier training to improve the efficiency and accuracy of annotation.
We implement a prototype system of MACPrint and evaluate it through extensive experiments.
The experimental results show that our system can efficiently and stably identify apps and actions performed by users, and speculate on user privacy.

\section{Threat Model}
\begin{figure}
    \centering
    \includegraphics[width=1.0\linewidth]{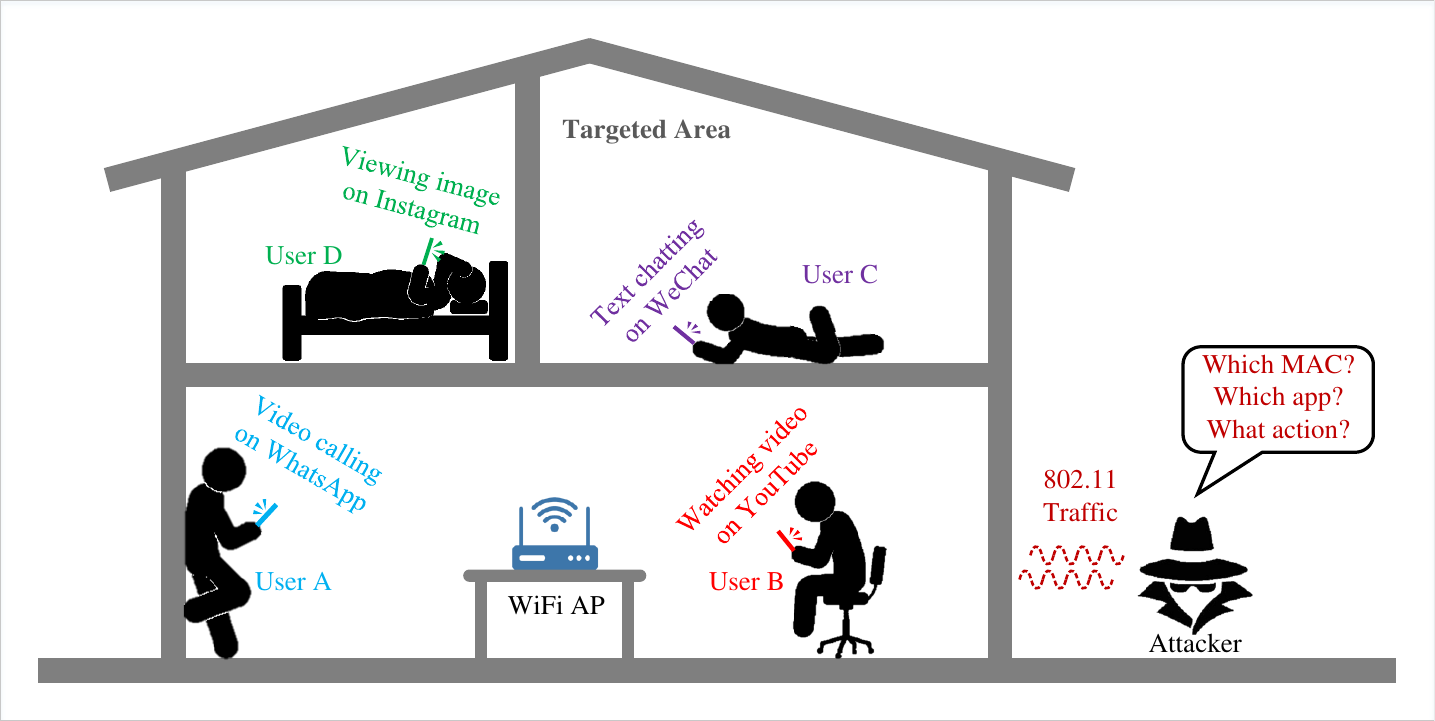}
    \caption{Threat model.}
    \label{fig:threatmodel}
\end{figure}

We consider a common scenario where a WiFi network is installed in an indoor area, such as a home or an office, to provide Internet access as depicted in Fig.~\ref{fig:threatmodel}. 
In this network, one access point (AP) can be deployed for sufficient wireless connectivity within the area. 
Moreover, multiple users could exist in this area, such as many family members living in a house.
Each user uses his smartphone, to join the network, and experience various app services, such as chatting on Twitter or watching videos on YouTube, based on his/her interests.

In this scenario, we consider traffic analysis attacks by observing the network's wireless encrypted traffic. 
Specifically, the adversary deploys sniffing tools within the reception range of the WiFi network but outside the targeted area in advance.
Since each AP would periodically transmit beacon frames with a consistent MAC address, the adversary can leverage Wireshark or Aircrack-ng to passively capture WiFi MAC frames from and to the targeted AP over the air.
His goal is to identify interested apps and actions from collected wireless traffic and thereby infer the relevant privacy of users.

\section{System Design}
\subsection{System Overview}
MACPrint is a novel system that identifies apps and actions via encrypted WiFi MAC frames.
MACPrint runs on a laptop or a desktop with a wireless network interface card and can be deployed around the targeted area, such as homes and offices, to sniff 802.11 traffic over the air.
It recognizes which app with which action on his/her smartphone, given sniffed wireless frames.
MACPrint exploits the side channel of WiFi communications to infer victims' privacy, making it completely passive and undetectable.
Furthermore, the user's age, gender, health status, sexual orientation, and other private information can be further inferred based on the identified apps and actions.

\begin{figure}
    \includegraphics[width=\textwidth]{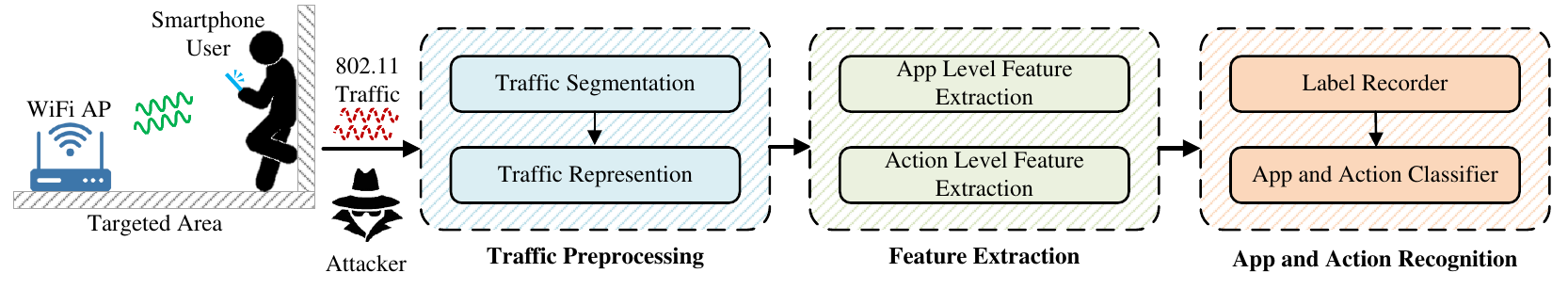}
    \caption{System overview.}
    \label{fig:SystemOverview}
\end{figure}

As shown in Fig.~\ref{fig:SystemOverview}, MACPrint consists of three core components. \textit{Traffic Preprocessing}, \textit{Feature Extraction}, and \textit{App and Action Recognition}.
\begin{itemize}
    \item \textbf{Traffic Preprocessing.} This component segments raw wireless traffic into effective traces and represents them using useful metadata encapsulated in MAC frames.
    \item\textbf{Feature Extraction.} To extract rich statistical and contextual features, we use sliding windows to extract multi-level features to characterize different apps and actions.
    \item \textbf{App and Action Recognition.} Our system exploits TCNs with OpenMax to recognize both used apps and actions in the open world setting. Additionally, an app called Label Recorder is developed to facilitate sample annotation for model training.
    % \item \textbf{User Profiling and Identification.} In this component, our system exploits the dynamic time warping algorithm to measure the distances of collected behavior sequences. 
    % Then, the silhouette coefficient method is used to determine the number of smartphone users. 
    % After that, our system relies on the k-means clustering for user profiling and identification.
\end{itemize}
\subsection{Traffic Preprocessing}
\textbf{Traffic Segmentation.} The first step of MACPrint is to sniff WiFi traffic from the targeted area and segment effective traffic traces that are generated by mobile users. 
To achieve this, MACPrint sets the wireless network card to the monitoring mode to scan all WiFi channels.
Once 802.11 frames are detected, it utilizes tools such as Wireshark and Aircrack-ng to capture them on the corresponding wireless channels. 
Due to background apps, smartphones can also generate traffic without user interaction. 
It is necessary to segment out traffic traces incurred by users. 
Generally, the traffic rate of background apps is much smaller, when compared with that of apps interacting with users~\cite{xiang2018appclassifier}. 
Therefore, traffic traces, i.e., sequences of consecutive frames, are picked out if the current packet rate is larger than a threshold $\gamma$. 
In our system, we empirically set $\gamma$ to be 3 frames per second.

\textbf{Traffic Representation.} After obtaining raw traffic traces, we proceed to filter out irrelevant 802.11 frames and extract useful metadata information from data frames. 
Typically, the collected wireless traffic contains three types of 802.11 frames, i.e., control frames, manage-ment frames, and data frames. 
The former two types carry little information about users' app usage behaviors. 
So we filter out management and control frames based on frame types and retain data frames only. 
Then extract the metadata information from the frame header to represent traffic traces.

% \begin{figure}
%     \centering
%     \includegraphics[width=0.9\linewidth]{FrameFormat.pdf}
%     \caption{802.11 Data frame format.}
%     \label{fig:frame_format}
% \end{figure}

Formally, we assume that a total of $T$ traffic traces are collected during the sniffing phase and $I$ MAC addresses are involved. 
Since users are not likely to use their smartphones all the time, multiple traffic traces may belong to the same address. 
Thus, the $j$-th traffic trace of $i$-th MAC address can be represented as
\begin{align}
    F^j_i = \left\{f^{1}, \cdots, f^m, \cdots, f^{M} \right\}.
\end{align}
In the above equation, $f^m=(t^m,s^m,d^m)$ represents the metadata of the $m$-th frame. 
Therein, $t^m\in \mathbb{R}$ and $s^m\in \mathbb{N}$ are, respectively, the arrival time and 
packet size of the $m$-th frame. 
$d^m \in  \left\lbrace -1,1 \right\rbrace $ represents downlink and uplink packets, which are transmitted from or to the AP, respectively.
$M \in \mathbb{N}$ stands for the total number of WiFi frames in the traffic trace $F^j_i$.
Moreover, let us denote $\Delta$ as the time duration of $F^j_i$. 
Hence, we have $\Delta=t^M-t^1$.
Finally, the collected traffic dataset $\mathcal{D}$ can be obtained as $\mathcal{D} = \left\{ F^j_i \right\}^{j=1:J_i}_{i=1:I}$, where $\sum_{i=1}^{I} J_i = T$.

\subsection{Multi-Level Feature Extraction}
Based on the captured traffic traces, we further extract multi-level features for recognizing apps and actions.
\begin{table}[htbp]
    \caption{Extracted In-App Action Features.}
    \resizebox{\linewidth}{!}{
    \begin{tabular}{|c|c|c|c|}
         \hline
         Tags & Features & Tags & Features\\
         \hline
         $p_1$& The total number of frames&$p_{12}$& Variance of uplink traffic(mid 60\%)\\
         $p_2$& Average size of all frames&$p_{13}$& Variance of uplink traffic(high 20\%)\\
         $p_3$& Average time interval&$p_{14}$& Average time interval for uplink traffic\\
         $p_4$& Uplink and downlink ratio&$p_{15}$& Average size of downlink traffic\\
         $p_5$& Kurtosis of frame rate&$p_{16}$& Average size of downlink traffic(low 20\%)\\
         $p_6$& Skewness of frame rate& $p_{17}$& Average size of downlink traffic(mid 60\%)\\
         $p_7$& Average size of uplink traffic& $p_{18}$& Average size of downlink traffic(high 20\%)\\
         $p_8$& Average size of uplink traffic(low 20\%)&$p_{19}$& Variance of downlink traffic(low 20\%)\\
         $p_9$& Average size of uplink traffic(mid 60\%)&$p_{20}$& Variance of downlink traffic(mid 60\%)\\
         $p_{10}$& Average size of uplink traffic(high 20\%)&$p_{21}$& Variance of downlink traffic(high 20\%)\\
         $p_{11}$& Variance of uplink traffic(low 20\%)&$p_{22}$& Average time interval for downlink traffic\\         
         \hline
    \end{tabular}
    }
    \label{tab:activity_feature}
\end{table}
Due to the differences in functionality and communication logic, different apps will present distinct changing patterns of the size, direction, and interval time among adjacent frames.
Since each traffic trace is a frame sequence, a feature extraction scheme using a sliding window is proposed to fully characterize the hidden sequential patterns.
The main idea is to divide each trace into equally sized feature samples with a sliding window. 
Specifically, centering at the $m$-th frame in the traffic trace $F^j_i$, the sliding window with a length of $W_s$ is used to generate an app feature sample $u^m$ as 
\begin{align}\label{eq:application_sample}
u^m = \left\{f^{m-(W_s-1)/2},\cdots,f^m,\cdots,f^{m+(W_s-1)/2}\right\}.
\end{align}
In this way, we can yield a feature sample for each frame in $F^j_i$.
From Eq.~\eqref{eq:application_sample}, we can observe that $u^m$ not only retains metadata information of the $m$-th frame but also preserves statistical and contextual features of its neighboring frames, which are sufficient for app recognition.

The next step is to extract fine-grained in-app action features.
In general, many fine-grained actions, such as text chat and browsing, are highly correlated with user actions, which are continuous in the time domain.
Therefore, we divide the traffic trace $F^j_i$ into namely traffic bursts, i.e., segments with a time duration of one second each.
In this way, a total of $N$ traffic bursts can be obtained, where $N\times1 \approx \Delta$, i.e., the time duration of $F^j_i$.
Given the $n$-th traffic burst, we extract 22 statistical features as listed in Table~\ref{tab:activity_feature} and obtain a feature vector $p^n=(p_1^n,\cdots,p_{22}^n) \in \mathbb{R}^{22}$.
Then, the sliding window with a size of $W_a$ is leveraged to generate in-app action feature samples.
Specifically, for the $n$-th traffic burst of $F^j_i$, the action feature sample $v^n$ can be written as 
\begin{align}
v^n = \left\{p^{n-(W_a - 1)/2},\cdots,p^n,\cdots,p^{n+(W_a - 1)/2}\right\}.
\end{align}
Through the above process, fine-grained statistical and contextual information can be extracted for action classification.

\subsection{App and Action Recognition}
Based on the multi-level features, we take the next step to recognize which apps of interest are used and which actions are performed by mobile users. 

\begin{figure}
    \centering
    \includegraphics[width=0.9\linewidth]{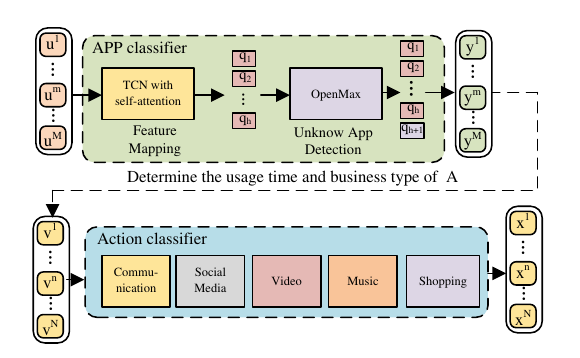}
    \caption{App classifier and action classifier.}
    \label{fig:Recognitionprocess}
\end{figure}
 
\textbf{App Classifier.} Taking an app feature sample $u^m$ as input, our app classifier $\mathcal{F}_{app}(\cdot)$ outputs a one-hot vector $y^m \in \left\{0,1\right\}^{H+1}$, where 1 indicates the corresponding app and $H$ is the number of apps of interest.  
In addition, the last element in $y^m$ stands for apps that are unseen in model training.
For effective recognition, the app classifier consists of feature mapping and unknown app detection as shown in Fig.~\ref{fig:Recognitionprocess}.

In feature mapping, we develop a temporal convolutional network with a self-attention mechanism to convert an app feature sample $u^m$ into a probability vector $ Q^m = [q^1,\cdots, q^{H}] \in [0,1]^{H} $.
The reasons for adopting TCNs are two-fold.
First, TCNs~\cite{bai2018empirical} are composed of a set of dilated causal convolutional layers and can abstract underlying temporal and contextual information from time series data, making them suitable for processing sequences of wireless frames.
Second, compared with traditional networks like RNNs, TCNs generally have better performance in training speed and scalability.
Moreover, we enhance our TCN model with a self-attention mechanism, which helps the model extract long-term dependencies in the input traffic trace.

In unknown app detection, we leverage an OpenMax~\cite{bendale2016towards} function to extend $H$-dimensional probability vector $ Q^m $ into a calibrated one $ \hat{Q}^m $ with $H+1$ dimensions to facilitate open-set classification.
Typically, most existing traffic fingerprinting techniques are based on closed-set classifiers, which can only handle known categories presented in model training but cannot correctly recognize unseen ones during the testing phase.
In practice, there are millions of apps in various mobile app stores, making it a non-trivial task to collect wireless traffic from all of them.
To deal with this issue, we employ an OpenMax function.
Mathematically, given the app feature sample $u^m$, this function converts the corresponding TCN output $ Q^m $ into $ \hat{Q}^m \in [0,1]^{H+1}$ as
\begin{align}\label{eq:calibrated_vector}
    \hat{Q}^m  = [q^1c^1,\cdots, q^hc^h,\cdots, q^{H} c^{H}, q^{H+1}],
\end{align}
where $q^{H+1}$ is the probability for unknown apps and can calculated by
\begin{align}\label{eq:probability_unknown}
   q^{H+1} = \sum_{h=1}^{H} q^n (1-c^h).
\end{align}
In the above two equations, $c^h \in [0,1]$ is a confidence weight, indicating the likelihood that $u^m$ belongs to the $h$-th category. 
The confidence $c^h$ is estimated by measuring the distance between $u^m$'s activation vector, i.e., the feature map of the last fully connected layer in the app classifier, and that of correctly classified training samples of the $h$-th category using the Weibull distribution.
In general, the closer they are, the bigger the $c^h$ is.
By using the OpenMax function, our app classifier can learn the feature space of known apps and detect unknown ones, thus improving its performance in open-world classification.
Using the calibrated probability vector $ \hat{Q}^m $, the app classifier predicts the most likely app $h^m$ as
\begin{align}
    h^m = \left\lbrace 
    \begin{array}{lcl}
    H+1 \: , & \text{if} \: \: q^{H+1} > \delta ; \\
    \arg \underset{{h=1:H}}{\max} \: \hat{Q}^m, \:  & \text{else} .
\end{array}
\right.
\end{align}
Therein, $\delta$ is a threshold. 
When $q^{H+1}>\delta$, the classifier predicts the sample $u^m$ as an unknown class. Otherwise, it can be considered from the app with the highest probability. 
Finally, we can utilize one-hot encoding to transform $h^m$ into the one-hot vector $y^m$.

Toward this end, we can feed all app feature samples in the traffic trace $F^j_i$ into our app classifier $\mathcal{F}_{app}(\cdot)$ and obtain a sequence of app labels as
\begin{align}
    Y^{j}_{i}  = \mathcal{F}_{app}(F^j_i) = \left\{y^1, \cdots, y^m, \cdots, y^M \right\}.
\end{align}
Then, to assign an app prediction to each traffic burst, we divide $Y^{j}_{i}$ into one-second segments based on burst timestamps.
For each segment, the most frequent label is selected as the app label of the corresponding burst.
In this way, we obtain another app label sequence $\hat{Y}^{j}_{i} $ as
\begin{align}
    \hat{Y}^{j}_{i}  = \left\{\hat{y}^1, \cdots, \hat{y}^n, \cdots, \hat{y}^N \right\}.
\end{align}
In addition, to effectively train our app classifier $\mathcal{F}_{app}(\cdot)$, we use the cross-entropy loss to measure the differences between its predictions and the ground-truth labels in the training phase.  

\begin{table}
    \centering
    \caption{40 Selected Apps From the China Android Market and Google Play Store.}
    \setlength{\tabcolsep}{0.3mm}{
    \begin{tabular}{|c|c|c|}
         \hline
         \textbf{App Category}& \textbf{Apps of Interest} & \textbf{In-App Actions}\\
         \hline
         \textbf{Messaging}&  
         \begin{tabular}{cccc}
              WeChat&QQ&WhatsApp&Telegram\\
              Messenger&Snapchat&Hangouts&Discord\\
         \end{tabular}
         &
         \begin{tabular}{cc}
              Text&Voice Chat\\
              Images&Video Chat\\
         \end{tabular}\\
         \hline
         \textbf{Social Media}& 
         \begin{tabular}{cccc}
              Weibo&Baidu Tieba&Quora&Facebook\\
              Twitter&Red Booklet&Reddit&Instagram\\
         \end{tabular}
         & 
         \begin{tabular}{cc}
              Browsing&Comment\\
              Thumb-up&Share\\
         \end{tabular}\\
         \hline
         \textbf{Video}&
         \begin{tabular}{cccc}
              YouTube&Tiktok&Netflix&Vimeo\\
              Tencent Video&Bilibili&Twitch&iQIYI\\
         \end{tabular}
         & 
         \begin{tabular}{cc}
              Forward&Play\\
              Backward&Next\\
         \end{tabular}\\
         \hline
         \textbf{Music}& 
         \begin{tabular}{c}
              \begin{tabular}{ccc}
              NetEase Cloud&QQ Music&Spotify\\
              \end{tabular}\\
              \begin{tabular}{ccc}
              SoundCloud&Apple Music&Shazam\\
              \end{tabular}\\
              \begin{tabular}{cc}
                  Kugou Music& YouTube Music\\
              \end{tabular}              
         \end{tabular}
         & 
         \begin{tabular}{cc}
              Forward&Play\\
              Backward&Next\\
         \end{tabular}\\
         \hline
         \textbf{Shopping}& 
         \begin{tabular}{cccc}
              Taobao&JD&PDD&Amazon\\
              eBay&Walmart&Rakuten&Suning\\
         \end{tabular}
         & 
         \begin{tabular}{cc}
              Search&Browsing\\
              Cart&Checkout
        \end{tabular}\\
        \hline
    \end{tabular}
    \label{tab:applist}
    }
\end{table}

\textbf{Action Classifiers.} As aforementioned, millions of mobile apps could run on smartphones, making it an impossible task to build a dedicated action classifier for each app.
For this reason, we select 40 high-ranking mobile apps from both the China Android Market and Google Play as the apps of interest as presented in Table~\ref{tab:applist}.
Then, based on their functions and services, we classify them into five categories and summarize the common actions of each category.

In this way, five action classifiers are built for each category. 
Without loss of generality, let us denote an action classifier as $\mathcal{F}_{act}(\cdot)$.
Given the action feature sample $v^n$, it outputs a one-hot vector $x^n \in \left\{0,1\right\}^{G+1}$, where $G$ indicates the total number of actions in all apps.
For simplicity, $\mathcal{F}_{act}(\cdot)$ has the same architecture with $\mathcal{F}_{a}(\cdot)$.
In the inference stage, MACPrint decides to use which action classifier based on predicted app labels.
Finally, given the traffic trace $F^j_i$, MACPrint outputs a sequence of action labels as
\begin{align}
    X^{j}_{i}  = \mathcal{F}_{act}(F^j_i) = \left\{x^1, \cdots, x^n, \cdots, x^N \right\}.
\end{align}

To this end, according to $\hat{Y}^{j}_{i}$ and $X^{j}_{i}$, we can obtain a user behavior sequence $B_i^j$ from the traffic trace $F^j_i$ as
\begin{align}
    B_i^j = \left\{b^1,\cdots,b^n,\cdots,b^N\right\}
\end{align}
where $b^n=\left\{y^n, x^n \right\}\in \left\{0,1\right\}^{H+G+2}$ is the concatenation of $y^n$ and $x^n$, and it represents the behavior of the $n$-th traffic burst, which uniquely indicates which app is used and which action is performed.

\textbf{Label Recorder.} Considering that our app classifier outputs a prediction for each MAC frame, a da-taset with frame-level annotation is necessary for model training. 
However, existing approaches mainly rely on manual annotation for each traffic trace, which cannot accurately label frames due to the frequent switch among different apps and actions by users in reality. 
To solve the problem, we design the Label Recorder, an Android app that collects log information relative to user interaction events. 
When running on the smartphone, Label Recorder captures log information about the foreground app and saves it as an interaction log in a txt format.
% More specifically, the label recorder works based on the TouchEventAccessibilityService class, where the currentTimeMillis, getPackageName, and getSource methods are leveraged to acquire the screen tapping time, the name of the foreground app, and the screen tapping location.
In this way, the label recorder allows us to obtain the interaction log as
\begin{align}
    L = \left\{ \cdots,l^{r-1},l^r,l^{r+1},\cdots \right\},
\end{align}
where $l^r=(t^r,b^r,d^r)$ represents the $r$-th records. 
Here, $t^r$, $b^r$, and $d^r$ denote the tapping time, the name of the foreground app, and the tapping location, respectively. 

During data collection, we run the label recorder on the user's smartphone to gen-erate the interaction log and exploit WiFi sniffing tools to capture corresponding traf-fic traces. 
To annotate traffic with specific apps, we exploit the temporal correlation between user interaction and the interaction log. 
To do this, we rely on the time and app name records to segment the period of the log into many intervals, during which the same app runs in the foreground. 
Subsequently, the traffic frame within the same interval is labeled with the corresponding app name. 
To annotate traffic traces with action labels, we leverage the spatial correlation between actions and the app user interface (UI) interface. Typically, each app has a fixed UI, and actions are activated by the click of some pre-defined locations on the screen. 
This implies that there is a correlation between the tapping location and actions, allowing us to infer performed actions from tapping location records. 
Based on this observation, we first establish a table about the correlation between the click location and actions for each app of interest. Subsequently, we label one traffic burst with a specific action if its correlated location is pressed within the burst time. 
Moreover, if no targeted action is detected, the burst is labeled with an unknown class. 
In this way, we can obtain labeled traffic traces to train our app and action classifiers.

Note that the label recorder is installed only on the controlled users' devices for training the app and action classifiers.
Once trained, they are deployed to infer apps and actions from the victim users' devices.

\section{Evaluation}
\subsection{Evaluation Methodology}

\textbf{Implementation.} We implement MACPrint using a Lenovo laptop connecting to a Kali dual-band net-work card. 
The laptop has an Intel Core i7-12700H processor and runs on the Kali Linux operating system. 
The working frequency range of the network card is 2.4~GHz-2.4835~GHz and 5.125~GHz-5.825~GHz. 
We set the network card to monitoring mode using Aircrack-ng to search all active WiFi APs. 
Then, we use Wireshark to collect MAC layer traffic from the targeted AP.

\textbf{Data Collection.} In our experiments, we build two datasets for performance evaluation. 
First, we con-struct Dataset A to evaluate MACPrint's performance in the closed world. 
During traffic collection, we first run our designed label recorder on the MI smartphone to record the user-device interaction log. 
Then, one user randomly performs defined and undefined actions on apps of interest to generate WiFi traffic. 
Next, we sniff the cor-responding wireless traffic from the MI smartphone using the Kali network card. 
In this condition, we collect 15 traffic instances for each app, with each instance lasting 10 minutes, and obtain traffic data over 100 hours. 
Based on the proposed traffic preprocessing, about 360000 samples are contained in Dataset A. 
Furthermore, we construct Dataset B to evaluate MACPrint's performance in the open world.
In addition to the apps included in Table~\ref{tab:applist}, we collect traffic instances of 10 apps such as Youku and Zhihu in the same way.
Therefore, a total of 450000 traffic samples are included in Dataset B.

\subsection{Experimental Results}

\textbf{Overall Performance.} First, we present the overall performance of MACPrint. 
The results indicate that MACPrint achieves an accuracy of over 96\% and an F1 score of over 96\% for rec-ognizing mobile apps and actions in the closed-world setting. 
When comes to the open-world app and action classification, our system obtains an accuracy of over 86\% and an F1 score of over 85\% in the open-world setting.

\begin{figure}
    \centering
    \includegraphics[width=0.75\textwidth]{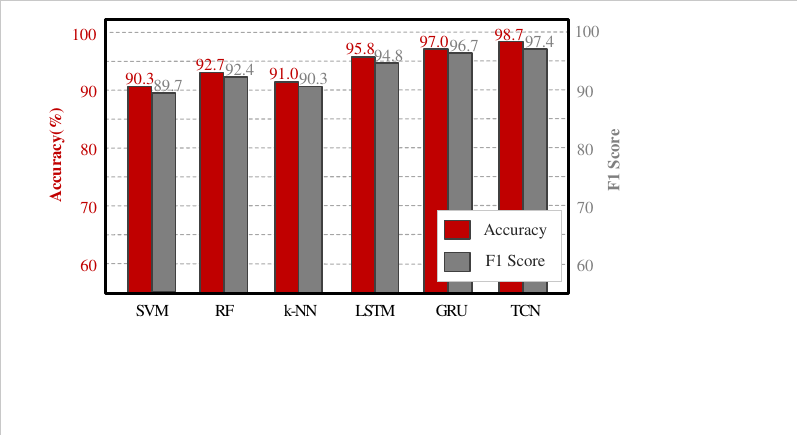}
    \caption{Performance of different classifiers.}
    \label{diffclassifiers}
\end{figure}

\textbf{Impact of TCN.} Then, we compare the adopted TCN with other classification models. 
To achieve this goal, we build five models and evaluate them on Dataset A. Specifically, our base-lines include GRU, LSTM, SVM, RF, and k-NN. 
In the GRU and LSTM, we set the number of hidden layers to 2, the layer size to 256, and the learning rate to 0.001.
Moreover, we choose the radial basis function as the SVM kernel. 
We set the number of estimators to 100 in the RF model and configure the number of nearest neighbors to 10 in the k-NN model. 
The experimental results are depicted in Fig.~\ref{diffclassifiers}.  
We can observe that the average accuracy and F1 score of all models reach 90\%. 
Notably, TCN, GRU, and LSTM achieve a classification accuracy of more than 97\% and outperform the other three models, indicating that deep learning models are better at app classification based on wireless traffic.
Among them, the TCN has the highest accuracy rate of 98.7\% and the best F1-score of 97.4\%.

\begin{figure}
    \centering
    \includegraphics[width=1\linewidth]{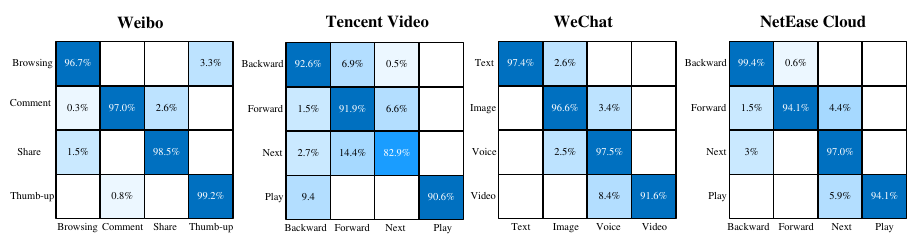}
    \caption{Confusion matrix of action classification on WeChat, NetEase CloudMusic, Tencent Video, and Weibo.}
    \label{fig:differentaction}
\end{figure}

Furthermore, we present the TCN's performance on action classification.
Fig.~\ref{fig:differentaction} showcases MACPrint's action classification performance on WeChat, NetEase CloudMusic, Tencent Video, and Weibo.
Through the confusion matrix, we can find that when using the trained TCN neural network, 
except for Tencent Video's four types of actions, the recognition accuracy of the other three apps' actions is above 94\%.
Even though the recognition accuracy of Tencent Video's related activities is low, it still reaches an accuracy of 90\%.
We suspect that the reason for the low accuracy in identifying video actions is that the four common actions all require a large amount of downlink traffic, which makes the traffic characteristics of different actions easily confused.
But our action classifier can still identify them relatively stably.

\textbf{Impact of Sliding Window Size.} Next, we show the impact of sliding window size on multi-level feature extraction. 
As mentioned in Section 3, we adopt a sliding window to extract app-level and action-level features. 
The choice of window sizes may impact the performance of MAC-Print. Theoretically, the larger the window size, the better performance is achieved. 
In our experiment, we vary the window size from 10 to 50 in the app classification. 
In each setting, app-level features are extracted from this window and fed into the app classifier. 
Similarly, we vary the window size from 2 to 10 when performing action classification.

\begin{figure}
    \centering
    \includegraphics[width=1\linewidth]{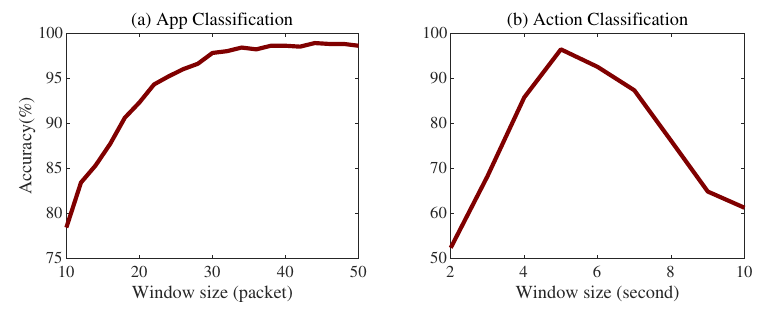}
    \caption{App and action classification based on different sliding windows.}
    \label{fig:differentwindow}
\end{figure}

As depicted in Fig.~\ref{fig:differentwindow}~(a), when the window size rises from 10 to 30, a rapid performance improvement can be observed.
However, when it goes beyond 30, the accuracy stays still around 98\%. 
This may be due to smaller windows containing fewer contextual features, while larg-er windows can provide richer app-level features, thereby improving classification accuracy. Based on the above result, we set the window size to 30 frames in app-level feature extraction.
As shown in Fig.~\ref{fig:differentwindow}~(b), an accuracy increase is observed when the window size increases from 2 to 5. 
But when it rises from 5 to 10, the classification accuracy decreases significantly. 
This is may due to that the majority of actions can be completed within 5s. When the window size exceeds 5s, traffic patterns from other actions could be introduced, lead-ing to an accuracy decrease. 
From the above observation, we set the window size to 5s in action-level feature extraction.

\begin{figure}
    \centering
    \includegraphics[width=0.9\linewidth]{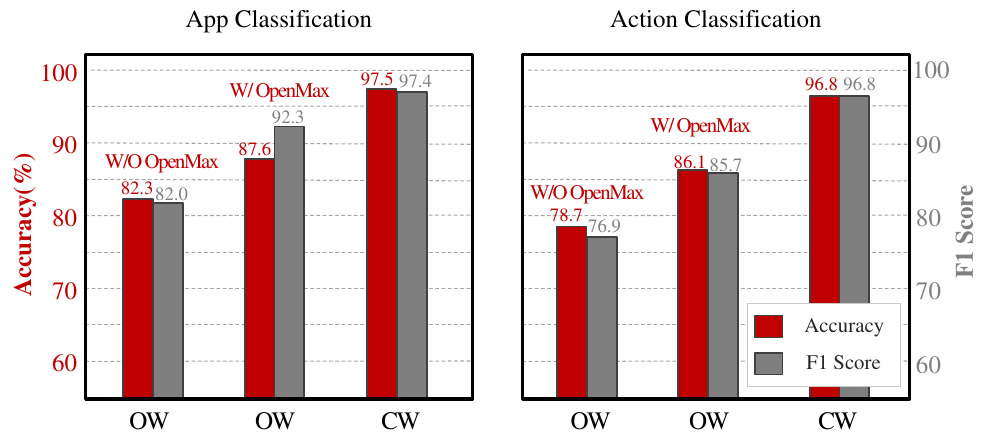}
    \caption{The impact of OpenMax.}
    \label{fig:openmax}
\end{figure}

\textbf{Impact of OpenMax.} Furthermore, we demonstrate the effectiveness of OpenMax functions for app and action classification in the open world.
For this purpose, we build a set of TCN classifiers that are without OpenMax functions and test them on Dataset B, which contains feature samples from both selected and unseen apps in Table~\ref{tab:applist}.
Fig.~\ref{fig:openmax} depicts the classification performance of MACPrint in different settings.
It can be observed that without the help of the OpenMax function, MACPrint obtains a low classification performance, where the app classifier has an accuracy of 82.3\% and an F1 score of 82.0\% and the action classifiers have an accuracy of 78.7\% and an F1 score of 76.9\%. 
The low classification performance is due to that traditional TCNs cannot deal with unseen samples, which are mistakenly classified into known categories in the testing phase. 
However, with the OpenMax function, the accuracy and F1 score of MACPrint increase to 87.6\% and 92.3\% in app classification, respec-tively, and are close to those in the closed-world setting. 
The performance boost owes to the ability of the OpenMax function to learn the distribution of known samples and discriminate unknown ones. 
Similar results can be found in the task of action classification. 
These observations manifest that the OpenMax function can effec-tively improve the overall performance of MACPrint in the open-world environment.

\section{Conclusion}
This paper proposes MACPrint, a novel traffic analysis attack system that can passively collect user traffic and identify the apps and actions used by the user by analyzing the encrypted traffic at the MAC layer.
Our system can effectively extract fine-grained features of wireless traffic to build app and action fingerprints, and can accurately identify apps and actions in the open world.
Finally, we implement MACPrint using a Lenovo laptop connecting to a Kali dual-band network card and evaluate it in the open world.
The experimental results show that our system can efficiently and stably identify apps and actions performed by users, and speculate on user privacy. 
This paper proposes and proves a reasonable and obvious threat to user privacy, and we hope that this study can enhance public awareness of privacy protection and promote detection research on upcoming attacks and new defense methods.

\bibliographystyle{splncs04}
\bibliography{ref}

\end{document}